\documentclass{emulateapj}
\usepackage{apjfonts}
\usepackage{graphicx}
\usepackage{amsmath}
\usepackage{natbib}
\usepackage{hyperref}
\newcommand{\msun}{\ensuremath{M_{\odot}}}

\newcommand{\lsun}{L{_{\odot}}}
\newcommand{\kms}{\ensuremath{\mathrm{km~s}^{-1}}}

\newcommand{\Hii}{\textrm{H}\,\textsc{ ii}}
\newcommand{\Cii}{c_{\textsc{ii}}}

\newcommand{\microns}{\mu \mathrm{m}}

\shorttitle{How to Find Young Massive Cluster Progenitors}
\shortauthors{E. Bressert et al.}

\begin{document}

\title{HOW TO FIND YOUNG MASSIVE CLUSTER PROGENITORS}

\author{Eli Bressert\altaffilmark{1,2},
 A. Ginsburg\altaffilmark{3}, J. Bally\altaffilmark{3},
 C. Battersby\altaffilmark{3}, S. Longmore\altaffilmark{1}, and L. Testi\altaffilmark{1,4}}

\altaffiltext{1}{European Southern Observatory, Karl Schwarzschild str. 2,
 85748 Garching bei M\" unchen, Germany}
\altaffiltext{2}{School of Physics, University of Exeter,
 Stocker Road, Exeter EX4 4QL, UK}
\altaffiltext{3}{Center for Astrophysics and Space Astronomy,
 University of Colorado, Boulder, CO 80309, USA}
\altaffiltext{4}{INAF-Osservatorio Astrofisico di Arcetri, 
 Largo E. Fermi 5, I-50125 Firenze, Italy}

\begin{abstract}
We propose that bound, young massive stellar clusters form from dense clouds that have escape speeds greater than the sound speed in photo-ionized gas. In these clumps, radiative feedback in the form of gas ionization is bottled up, enabling star formation to proceed to sufficiently high efficiency so that the resulting star cluster remains bound even after gas removal. We estimate the observable properties of the massive proto-clusters (MPCs) for existing Galactic plane surveys and suggest how they may be sought in recent and upcoming extragalactic observations. These surveys will potentially provide a significant sample of MPC candidates that will allow us to better understand extreme star-formation and massive cluster formation in the Local Universe.
\end{abstract}


\section{INTRODUCTION}

The formation of bound star clusters has become a topic of renewed interest. The Milky Way contains about 150 globular clusters (GCs) with masses from $10^4\ \msun$\ to over $10^6\ \msun$\ and tens of thousands of open clusters containing from 100 to over $10^4$ stars \citep{PortegiesZwart2010}. While no GCs have formed in the Milky Way within the last 5 Gyr, bound clusters that survive for more than hundreds of crossing times continue to form.

Infrared observations over the last two decades have shown that molecular clouds tend to produce stars in higher surface densities ($\geq 3~{\rm stars~pc}^{-2}$) than the field population \citep{Lada2003}. \cite{Bressert2010} showed that stars within 500 pc of the Sun form in a smooth continuous distribution and only a minority will dynamically evolve to form bound low-mass stellar clusters ($10^2$ to $10^3 \msun$). The vast majority of these young clusters are {\it transient} groups that are bound primarily by the gas in their environment. Thus, while most stars may form in groups, gravitationally bound clusters which remain bound for many crossing-times following dispersal of their natal clump are rare and contain less than 10\% of the Galactic stellar population. Despite the small number of stars that form in the bound young massive clusters \citep[YMCs; $\gtrsim 10^4$ \msun; ][]{PortegiesZwart2010}, they are important as they shed light on extreme star-formation in the Local Universe and provide insight on how GCs may have formed in the high-redshift universe and in the distant past of the Milky Way \citep{Elmegreen1997}.

With new Galactic plane surveys, e.g., the Herschel HiGAL survey \citep{Molinari2010}, the APEX Telescope Large Area Survey of the Galaxy (ATLASGAL; \citealt{Schuller2009}), the Bolocam Galactic Plane Survey (BGPS; \citealt{Aguirre2011}), the H$_2$O Southern Galactic Plane Survey (HOPS; \citealt{Walsh2011}), and the Millimeter Astronomy Legacy Team 90 GHz Survey (MALT90; \citealt{Foster2011}), we are on the cusp of better understanding how massive clusters form. The question is, how do we find the massive proto-clusters (MPCs) that will form these YMCs? We investigate how the YMCs may form and provide a simple model with observational properties that can be used to identify MPC candidates. To identify more extreme MPCs in nearby galaxies (e.g., the Antennae Galaxies), we need capable telescopes like the Atacama Large Millimeter/sub-millimeter Array (ALMA).

YMCs may predominantly form from MPCs having gravitational escape speeds greater than the sound speed in photo-ionized gas. When this condition is met, ionization cannot disrupt the entire MPC.  Stars can continue to form from the remaining neutral gas and star formation efficiency (SFE) increases to 30\% and higher. The remaining mass in the stellar population nullifies the effects of gas expulsion and the cluster will remain bound. If the absolute value of the gravitational potential energy is greater than the expected thermal energy of the plasma in a massive gas clump, and supernovae have not yet occurred, then we consider the object to be an MPC candidate. We compare the effective photo-ionized sound speed of the plasma ($\Cii$) to the escape velocity ($v_{esc}$) of the clump, which we denote as $\Omega \equiv v_{esc} / \Cii$. A gas clump that has $\Omega>1$, implying $v_{esc} > \Cii$, is an MPC candidate while $\Omega<1$ are not, since gas can be dispersed by the appearance of the first OB stars.

We describe the simple model of massive stellar cluster formation using the $\Omega$ parameter in \S\ \ref{sec:ymcf}. We make predictions on the MPC's observational properties for Galactic plane surveys and ALMA (extragalactic) in \S\ \ref{sec:predictions}. \S\ \ref{sec:conclusions} discusses the implications of the model and predictions with a summary of the results.

\section{YOUNG MASSIVE CLUSTER FORMATION}
\label{sec:ymcf}

\subsection{Initial Conditions}
What are the initial conditions that lead to bound stellar cluster formation, e.g., YMCs like NGC 3603 and R136? We must consider the differences between low-mass and high-mass star forming clouds. In the solar neighborhood the local SFE, defined as the final mass in stars formed in a cloud divided by the initial mass of gas, of the low-mass star forming regions is reported to be ~5\% or less \citep{Evans2009}. Numerical experiments \citep[e.g., ][]{Lada1984a,Geyer2001,Goodwin2006} demonstrate that the formation of gravitationally bound clusters requires that the local SFE $\sim$30\%. Observations show similar results of higher SFE for massive star forming regions\citep[e.g., ][]{Adams2000,Nurnberger2002} and hint that some of these systems could evolve into open clusters. The initial conditions that enable the formation of YMCs may be responsible for the high SFE observed in these bound stellar clusters. From the gas clump phase to the stellar clusters the gas is somehow efficiently converted to stars. Assuming that the SFE of 30\% is critical for forming a YMC, then the MPC mass should be greater than $3 \times 10^4\ \msun$.

Star formation in a clump ends when its gas is dispersed by the feedback energy injected from its newborn stars. Gas can be dispersed when the outward pressure generated by feedback exceeds the inward pressure of the overburden of gas in the cluster gravitational potential well \citep{Bally2011a}. The ratio of the local escape velocity divided by the sound speed for photo-ionized gas, $\Omega = v_{esc} / \Cii$, plays a crucial role in determining if a clump forms a bound cluster or not. Feedback from accreting low-mass protostars is dominated by bipolar outflows \citep{Bally2011a}. As stars reach several solar masses, their non-ionizing near-UV radiation photo-heats surrounding cloud surfaces, raising the sound speed to $\sim$ 1 to 5 \kms. Stars greater than 10 \msun\ (early-B and O stars) ionize their surroundings, raising the sound speed up to $\sim$ 10 $\kms$. Although protostellar outflows, stellar winds, and radiation pressure of ever-increasing strength can also raise the effective sound speed by generating internal motions, only the effects of ionizing radiation will be considered here. See \S\ 2.2 for details.

If $\Omega < 1$, the gas can be dispersed in a few crossing times from the star-forming clump, bringing star formation to a halt with a low stellar density. On the other hand, when $\Omega > 1$ the gas will remain bound and can continue to form new stars or accrete onto existing ones until further increases in stellar luminosity or mechanical energy injection raise the effective sound speed to a value greater than the escape speed. As the stellar mass increases, energy released by the forming embedded cluster grows \citep{Miesch1994}. In such a cluster, photo-ionized plasma will remain gravitationally bound by the cluster potential and recombinations in the ionized medium will tend to shield denser neutral clumps, allowing star formation to proceed to high efficiency. This model works under the assumption that the entire gas clump is instantly ionized, which is a worst case scenario for a clump to remain bound. This means that even if there is a large number of OB stars present in a clump, its gas will unlikely be fully ionized and our model would still hold. We discuss the details of ionization and its effect on pressure balance and ongoing star formation below.

\subsection{Ionizing Feedback, Pressure, and Star Formation}
\label{sec:feedback}
Feedback, which plays a major role in the self-regulation of star formation can be subdivided into two forms: {\it mechanical feedback} consisting of protostellar outflow, stellar winds, and supernova explosions and {\it radiative feedback} which can be subdivided into radiation pressure, non-ionizing FUV heating, and ionizing EUV heating. In massive cluster forming environments, prior to the explosion of the most massive star, stellar winds will dominate mechanical feedback and all three radiative mechanism can be important \citep{Bally2011a}.

Consider a fiducial reference clump of $M_{gas} = 2 \times 10^4 \msun$ and $M_{stellar} = \sim1 \times 10^4 \msun$ (SFE $=30\%$), a total luminosity $L_f$, Lyman continuum photon luminosity $Q_f$, and a total stellar wind mass loss rate $\dot M_f$ at $V_f$ (wind terminal velocity) which corresponds to 50 O7 stars (30 \msun) located at the center of the cluster. The assumed quantities for the fiducial cluster are shown in the denominators of the equations below \citep[see ][ and references therein]{Martins2005,Donati2002}. 
Through 10$^4$ Monte Carlo iterations the ratios of $L$, $Q$, and $\dot M$ between the stellar population of the fiducial cluster and Salpter-like cluster ($8\ \msun < M_{star} <110\ \msun$) are shown to be $L_f/L_{Salpeter} = 14.0$, $Q_f/Q_{Salpeter} = 2.2$, and $\dot M_f/\dot M_{Salpeter} = 8.9$. These calculations were done using \cite{Murray2010a} for Q and \cite{Crowther2000} for $\dot M$. The gravitational radius for this cluster $r_G = 2 G M / \Cii ^2$ = 2.13 pc and $\alpha_B \approx 2.6 \times 10^{-13} \textrm{cm}^3 \textrm{s}^{-1}$ is the case-B recombination rate for H at a temperature of $10^4$ K. 

Below, we consider the radiation pressure, stellar wind ram pressure, and the internal pressure of a uniform density $\Hii$ region. These pressures are evaluated at $D=2$ pc from the center of the cluster, a distance close to the gravitational radius for the reference cluster.

\begin{enumerate}
{\item Stellar wind:
\begin{equation}
\begin{split}
    P_{w} &= \rho(r) V^2 = \dot M V / {4 \pi D^2}\\
    &= 1.32 \times 10^{-10}\, \textrm{dynes cm}^{-2}\,
    \left[\tfrac{\dot M_f}{5 \times 10^{-6} \msun~yr^{-1}} \right]
    \left[\tfrac{V_f}{2 \times 10^3\ \kms} \right]
    \left[\tfrac{D_f}{\textrm{2 pc}} \right]^{-2}\\
\end{split}
\end{equation}}

{\item Radiation pressure on an optically thick surface:
\begin{equation}
\begin{split}
    P_{rad} & = L / 4 \pi c D^2\\
    & = 2.68 \times 10^{-9}\, \textrm{dynes cm}^{-2}\,
    \left[\tfrac{L_f}{10^7\ \lsun} \right]
    \left[\tfrac{D_f}{\textrm{2 pc}} \right]^{-2}\\
\end{split}
\end{equation}}

{\item Thermal pressure of a uniform density $\Hii$ region with
Str\" omgren radius equal to $D = 2$ pc:
\begin{equation}
\begin{split}
   P_{\Hii} &= \mu m_H \Cii^2  (3 Q / 4 \pi \alpha_B)^{1/2} D^{-3/2}\\
   &= 5.40 \times 10^{-9}\, \textrm{dynes cm}^{-2}\,
   \left[\tfrac{Q_f}{10^{51} \textrm{photons s}^{-1}}\right]^{1/2}
   \left[\tfrac{D_f}{\textrm{2 pc}} \right]^{-3/2}\\
\end{split}
\end{equation}}
\end{enumerate}
\noindent These pressures are listed in order of increasing significance for the fiducial reference cluster. It is important to note that the equations express the feedback pressures as power laws of distance from the center of the fiducial cluster. The pressure from Eq. 3 tapers off the slowest among the three pressures, which bolsters the dominance of photo-ionization at large distances. A similar result was found in \cite{Krumholz2009}, but \cite{Murray2010b} conclude that radiation pressure is the dominant mechanism in dispersing gas under different scaling assumptions. If radiation does play a role in dispersing the gas it could mass-overload the shell around the \Hii\ region and force some gas to escape, which may imply that a YMC's SFE is unlikely to ever be 100\%. However, the ionized gas pressure will not unbind the system and could create large optically thin bubbles, allowing much of the radiation pressure to escape without interacting with any gas particles.

\section{PREDICTING OBSERVED PROTO-CLUSTER PROPERTIES}
\label{sec:predictions}

\subsection{Proto-cluster Geometries}
What are the expected physical geometries of MPCs? Using both the YMC progenitor model and observations of YMCs, we derive the maximum sizes and mass of the MPCs. In the bound $\Hii$ gas clump model, the key requirement is $v_{esc}>\Cii$ to form a YMC. By setting the potential and kinetic variables to $-G M m r^{-1}$ and $\frac{1}2 m \Cii^2$, we can solve for $r$. We denote this radius as $r_{\Omega}$, which is a function of the clump mass, $r_{\Omega}(M_{clump}) = 2 G M_{clump} \Cii^{-2}$. For MPC candidates with masses of $3 \times 10^4 - 3 \times 10^6$\ \msun, $r_{\Omega}$ ranges between 2.6 and 258.1 pc (see Table 1 for details). The $r_{\Omega}$ values for $>10^6\ \msun$\ clumps are large and are very unlikely to form a $>10^6 $\ \msun\ gravitationally bound star cluster as the free-fall time of the system could exceed when supernovae could begin and disrupt the system. Hence, an upper size limit for these objects is needed.

There is a well-measured constraint on YMCs that we can apply to predict the \textit{upper limit radii} for the MPCs. Recent high resolution imagining and spectral studies of YMCs like NGC 3603, Arches, Westerlund 1, and R136 have shown these systems to be in or close to virial equilibrium at ages of $\sim 1$ Myr \citep{Rochau2010,Clarkson2011,Cottaar2012,HenaultBrunet2012}. This implies that the YMCs must have gone through at least one full crossing time before their presently observed age. Using the crossing time equation from \cite{PortegiesZwart2010}, $t_{cross} = (G M r_{vir}^{-3})^{-1/2}$, and fixing the crossing time to 1 Myr we can solve for $r_{vir} = (G M t_{cross}^2)^{1/3}$, which is only dependent on the mass. For the same clump mass range mentioned for $r_{\Omega}$ above, $r_{vir}$ spans from 5.1 to 23.8 pc (see Table 1 for details). Note that the $r_{\Omega}$ and $r_{vir}$ values are the upper limit radii for the MPCs. This is an important aspect to keep in mind as there is no evidence for YMCs to have a proportionality between mass and radius \citep{Larsen2004,Bastian2012}. $r_{\Omega}$ and $r_{vir}$ have similar radii between $10^4 \msun$\ and $10^5 \msun$\, intersecting at $8.4 \times 10^4 \msun$, but at $>10^6\ \msun$\ $r_{\Omega} \gg r_{vir}$. Extragalactic predictions regarding more massive MPCs ($3 \times 10^7$ - $3 \times 10^9$ \msun) is provided in Table 1.

Figure 1 shows the upper limit radii relative to clump masses for $r_{\Omega}$ and $r_{vir}$. We include a shaded region marking where MPC candidates reside. Infrared dark clouds and clumps reported in \cite{Rathborne2006} and \cite{Walsh2011} are included to show where they lay on the plot relative to $r_{\Omega}$\ and $r_{vir}$. Five gas clumps fit within the MPC criterion: G0.253+0.016 \citep{Longmore2012}, an MPC candidate in the Antennae galaxy \citep{Herrera2012}, and three Galactic clumps reported in \cite{Ginsburg2012} which are discussed in further detail in their paper. 

We took a collection of Galactic YMCs summarized in \cite{PortegiesZwart2010} and estimated their progenitor masses by multiplying their current stellar mass by a factor of 3 to account for SFE$\sim30\%$. Eight of the of 12 \textit{estimated} YMC progenitors are within the virial and $\Omega$ radii limits. The other four are found not only close to the upper limit of the $r_{\Omega}$ line, but appear to follow the line. If we assume that the SFE $\gtrsim30\%$ then the estimated YMC progenitors will move above $r_{\Omega}$ line.

\subsection{Prediction and Observations}
From the derived physical properties of the MPCs discussed, we predict the MPC's integrated fluxes in the HiGAL, ATLASGAL, and BGPS surveys in Table 1 using mass to flux conversions discussed in \cite{Kauffmann2008} assuming T = 20 K. If T = 40 K, then the flux will increase by a factor of $\sim2$. Between the wavelengths of 500 $\microns$ and 1100 $\microns$ the emission from the bright MPCs and the interstellar medium are optically thin throughout the Galactic plane. \textit{We predict that the surveys should be sensitive to all of the MPCs in the Milky Way.} \cite{Tackenberg2012} independently came to similar conclusions for ATLASGAL. Moreover, the sources for the mass clumps $\geq 3 \times 10^4\ \msun$ are resolvable up to 20 kpc. We calculated the clump central pressures from \cite{Johnstone2000}. 
These values are provided in Table 1. Note that HiGAL has advantages over ATLASGAL and BGPS, since it has no spatial filtering and has the highest sensitivity amongst these surveys. 

\cite{Ginsburg2012} used the predicted properties to develop a selection criteria from the BGPS catalog and discovered three MPC candidates. The masses of the objects can be determined from the dust emission ($M_{dust}$), but comparing the dust masses to virial mass, $M_{vir}$, is important in determining whether a MPC candidate is gravitationally bound or not. Following similar treatment from \cite{Longmore2012} one can use $\Delta V$ from molecular line tracers (e.g., $\mathrm{HCO+}$ and $\mathrm{N_2H+}$) to approximate a clump's virial mass, $M_{vir} \propto r_{clump} \Delta V^2$. If $M_{dust} \sim M_{vir}$, then we can assume that the clump is consistent with being gravitationally bound. Estimating the mass of a clump from continuum emission requires a known distance to the objects using $V_{LSR}$, which line surveys can provide \citep[e.g., ][]{Walsh2011,Schlingman2011}. The combination of these line and continuum Galactic plane surveys will allow us to obtain a near complete census of MPCs in the Milky Way.

In Table 1, we show our predictions for the most massive MPCs ($>10^7$\ \msun) in the Local Universe regarding ALMA's observing capabilities. The given fluxes are calculated for the MPCs at a distance of 40 Mpc and $\theta \sim 0''.5$. These objects, if near the upper limit of $r_{vir}$, could be resolvable at these distances. 

\begin{figure*}
\begin{center}
\includegraphics[width=6.5in]{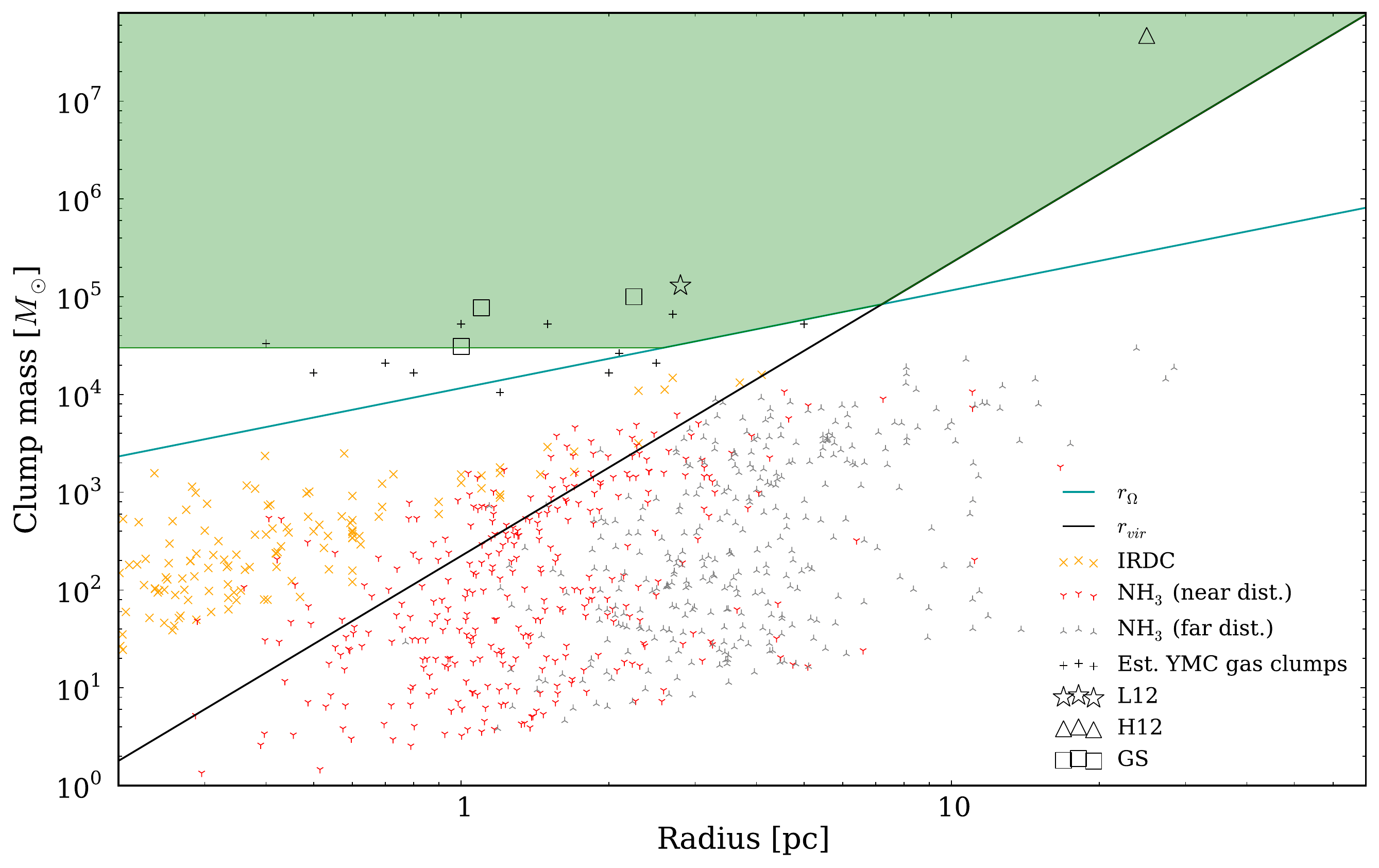}
\caption{
\label{fig:fig01}
The mass-radius parameter-space for clumps partitioned by radii for $r_{\Omega}$ (solid blue) and $r_{vir}$ (solid black). MPC candidates are defined with the following properties (green shaded region): a minimum mass of $3 \times 10^4 \msun$, $r > r_{\Omega}$\ for $M_{clump} < 8.4 \times 10^4 \msun$, and $r > r_{vir}$\ for $M_{clump} > 8.4 \times 10^4 \msun$. Clump masses and sizes are plotted on top from three different data catalogs: IRDCs \citep{Rathborne2006}, HOPS clumps \citep{Walsh2011}, and YMCs \citep{PortegiesZwart2010}. The YMCs are converted to their possible clump progenitors by assuming that SFE is $\sim$30\%, which boosts the mass of the systems by a factor of $10/3$. The scaled YMC progenitors happen to lie near the critical $r_{\Omega}$ line without any tweaking of parameters. Two published sources that have radii less than both their respective $r_{\Omega}$ and $r_{vir}$ are G0.253+0.016 \citep[L12; ][]{Longmore2012} and an {\it extragalactic} massive proto-cluster candidate reported in Herrera et al. (H12; 2012). The MPC candidates reported in Ginsburg et al. (GS; 2012) are shown as squares.}
\end{center}
\end{figure*}

\begin{table*}
\begin{center}
\caption{\label{tab:properties}
Predicted Proto-cluster Properties}
\begin{tabular}{cccccccc}
\hline
\noalign{\smallskip}&&Galactic&$d \leq 20$\ kpc&&\\
\noalign{\smallskip}\hline
Mass & $\log P_{\textrm{cen}} / k$ & $r_{\Omega}$ & $r_{vir}$ & $\theta_{\textrm{20 kpc}}$ & $F_{500}$ & $F_{850}$ & $F_{1100}$\\
$\msun$ & K/m$^2$ & pc & pc & '' & Jy & Jy  & Jy \\
\hline\hline
$3 \times 10^4$ &  13.15 & 2.6   & 5.1  &  53.62 & 30.78 & 4.17 & 1.81\\
$3 \times 10^5$ &  13.81 & 25.8  & 11.0 & 226.89 & 17.19 & 2.33 & 1.01\\
$3 \times 10^6$ &  14.47 & 258.1 & 23.8 & 490.91 & 36.72 & 4.98 & 2.16\\

\hline
\noalign{\smallskip}&&Extragalactic&$d \leq 40$\ Mpc&$\theta_{ALMA} = 0''.5$&&\\
\noalign{\smallskip}\hline
Mass & $\log P_{\textrm{cen}} / k$ & $r_{\Omega}$ & $r_{vir}$ & $\theta_{\textrm{40 Mpc}}$ & $F_{450}$ & $F_{850}$ & $F_{1200}$\\
$\msun$ & K/m$^2$ & pc & pc & '' & mJy & mJy  & mJy \\
\hline\hline
$3 \times 10^7$ &  15.13 & - & 51.2  & 0.53 &  77.7 &  7.8 &  2.3\\
$3 \times 10^8$ &  15.80 & - & 110.3 & 1.14 & 167.5 & 16.7 &  5.1\\
$3 \times 10^9$ &  16.47 & - & 237.7 & 2.45 & 360.6 & 36.6 & 11.0\\
\hline

\end{tabular}
\end{center}
\scriptsize
\textit{Top half:} The predicted physical properties of the YMC progenitors, which are observable throughout the Milky Way ($<=$20 Kpc) with the Hi-GAL, ATLASGAL and BOLOCAM Galactic plane surveys. \textit{Bottom half:} the predicted physical properties of the YMC progenitors, which are observable up to 40 Mpc with ALMA assuming an angular resolution of 0.5''. The values for $r_{\Omega}$ are not shown for the extragalactic sources since $r_{vir} \ll r_{\Omega}$ in all cases and hence such objects are immediately considered as PMC candidates. All fluxes are derived using $r_{\Omega}$ for the $\leq 8.4 \times 10^4\ \msun$\ and $r_{vir}$ for $> 8.4 \times 10^4\ \msun$. Note that the fluxes for the extragalactic clumps are calculated using the rest wavelengths. 
\end{table*}

\section{DISCUSSION AND SUMMARY}
\label{sec:conclusions}

We have discussed the possible conditions necessary for YMC formation and how to identify their progenitors, MPCs, in their primordial state regarding their masses, radii, and flux brightness. The key to identifying whether a massive gas clump can form a YMC is the balance between the gravitational potential of the gas clump and the gas kinematics. We characterize this balance as the ratio between $v_{esc}$ at a given radius and the sound speed of the photo-ionized gas, $\Cii \sim 10 \kms$. If $\Omega = v_{esc} / \Cii > 1$ (equivalent to, $v_{esc} > \Cii$) for a clump of gas ($>3 \times 10^4 \msun$) then the system will optimally convert the clump to stellar mass and likely form a YMC. We classify such clumps as MPC candidates. If $\Omega<1$ then the system does not have a deep enough potential well to keep the photo-ionized gas bound, which will lead to rapid gas dispersal and low star formation efficiency. The end product will be a low mass cluster or group that will feed its stars to the field star population over a short time scale. It may be possible that some of these MPC candidates will not form YMCs due to the cruel cradle effect, where the forming cluster is disrupted by nearby massive GMCs \citep{Kruijssen2011}.

With the HiGAL, ATLASGAL, and BGPS Galactic plane surveys we should be sensitive to the MPCs throughout the Milky Way. The combination of these surveys will provide us a near complete sample of the MPCs in the Milky Way and pave a path for future high-resolution studies. Furthermore, with ALMA's full potential we should be sensitive to MPCs over $>10^7$\ \msun\ in extragalactic systems within 40 Mpc of the Milky Way. This would help us better connect the Galactic and extragalactic MPCs to better understand the formation and evolution of YMCs, some of which could produce long lived ``young globular clusters'' \citep[see][and references therein]{PortegiesZwart2010}. It is important to note that $r_{vir}$ is the upper limit to these very massive proto-clusters and such objects will likely not be close to such scales. Observing these extragalactic MPC candidates in the nearby galaxies will help constrain the upper limit radii. 


\acknowledgments{This work was supported by NSF grant AST0407356. The authors would like to thank the referee, N. Bastian, D. Kruijssen, M. Krumholz, M. Bate, J. Dale, D. Johnstone, M. Gieles for constructive feedback and suggestions.}


\end{document}